\documentstyle[aps,pra,epsf]{revtex} 
\begin{document}
\draft
\tighten
\onecolumn
\title{Gate errors in solid state quantum computer architectures}
\author{Xuedong Hu and S. Das Sarma}
\address{Condensed Matter Theory Center, Department of Physics,
University of Maryland, College Park, MD 20742-4111}
\date{\today}
\maketitle
\begin{abstract}
We theoretically consider possible errors in solid state quantum
computation due to the interplay of the complex solid state
environment and gate imperfections.  In particular, we study two
examples of gate operations in the opposite ends of the gate speed
spectrum, an adiabatic gate operation in electron-spin-based quantum
dot quantum computation and a sudden gate operation in Cooper pair
box superconducting quantum computation.  We evaluate quantitatively
the non-adiabatic operation of a two-qubit gate in a
two-electron double quantum dot.  We also analyze
the non-sudden pulse gate in a Cooper-pair-box-based quantum computer
model.  In both cases our numerical results show strong influences of
the higher excited states of the system on the gate operation,
clearly demonstrating the importance of a detailed understanding of
the relevant Hilbert space structure on the quantum computer
operations. 
\end{abstract}
\pacs{PACS numbers:
03.67.Lx, 
85.25.Cp 
85.30.Vw, 
}

The rapid progress of quantum information theory has prompted 
intense interest in developing appropriate hardwares for quantum
computation and communication \cite{Review}.  Since quantum
coherence needs to be maintained for a quantum computer (QC) to
maximize its capability, it has to satisfy stringent requirements
\cite{DiVincenzo}.  One requirement is that a QC should have a
well-delineated Hilbert space, while another requirement is that the
quantum bits (qubit) should have long coherence time (i.e. slow
decoherence).  Indeed, decoherence can be loosely defined as
irreversibly losing information from the operational Hilbert space
into the environment, which is the rest of the universe from the
perspective of a QC system.  Furthermore, the requirement of slow
decoherence needs to be satisfied not only when the qubits are
standing alone, but also when the qubits are interacting with each
other or are manipulated externally for single or multi-qubit
operations.  

Many physical systems with suitable two-level quantum dynamics have
been suggested as possible candidates for qubits in a variety of QC
architectures.  The proposals based on solid state systems, such as
the ones based on electron spins \cite{LD,HD,HD2,Vrijen}, nuclear
spins \cite{Kane,Privman}, and superconductors \cite{Schon,Averin},
are particularly attractive because of their potential for
scalability.  However, solid state structures also present
complex environments and  fast decoherence rates \cite{HSD}.  For
example, for electron spins trapped in quantum dots (QD), their
environment includes the underlying crystal lattice in terms of
phonons, surrounding electron and nuclear spins, impurities and
interfaces, unintentionally trapped charges, and many other defects
invariably present in a solid state environment. 
Even the orbital degrees of freedom of the electron spin qubits
themselves are part of the environment.  In the case of QC based on
Cooper pair boxes (CPB), the environment includes defects in the
Josephson junctions, quasiparticles, electromagnetic couplings, etc.
From the perspective of qubit Hilbert space, the complex
environment in a solid means considerable difficulty in identifying
the boundary.  The fuzziness in the qubit environment also implies
that a gate operation that is not optimized (a gate error) can cause
not only incomplete rotations in the qubit Hilbert space, but also
leakage into the environment.  Thus, to ensure successful operation
of a QC, we have to clearly understand how each of the environmental
elements affects the QC coherent evolution, and clarify the effects
of imperfections in gate operations.  In this paper we will
investigate specifically the latter effect.

Some of the QC proposals satisfy the requirement for precise
coherent manipulation by employing resonant means such as resonant
Rabi oscillation and resonant Raman scattering \cite{iontrap}. 
However, in most solid state QC schemes, non-resonant operations
\cite{LD,Kane,Schon} are crucial components of the QC operation.  In
the spin-based schemes, adiabatic switching of exchange interaction
is a cornerstone of the two-qubit operations, while in the CPB-based
scheme, sudden switching of a voltage bias has been proposed and used
to  perform single-qubit operations.  Here we study whether QC system
integrity can be faithfully maintained if the exchange interaction
switching is not adiabatic in the QD case, or the voltage pulse is
not switched on suddenly in the CPB case.  More specifically, we would
like to explore the gate operation threshold when leakage due to
non-resonant operations is sufficiently small so that quantum error
correction codes can be effectively applied.  A quantitative and
qualitative theoretical investigation of this issue is clearly
important in understanding the feasibility of proposed QC
architecture.

Let us first examine the spin-based quantum dot quantum computer
(QDQC) in GaAs.  Here effects of non-adiabaticity in the single-qubit
operations should be relatively weak because bulk spin-orbit coupling
at the bottom of GaAs conduction band is small.  If the Rashba
spin-orbit coupling can also be minimized for the GaAs
heterostructure, the overall spin-orbit coupling should be very small
so that applied magnetic field pulses will not be able to mix
electron spin and orbital states.  For two-qubit
operations, the adiabatic condition can be satisfied more easily with
larger single electron excitation energy and larger on-site Coulomb
interaction, which are both consequences of small QDs.  However, the
size of a gated quantum dot is limited from below by gate and device
dimensions.  Thus we would like to quantitatively assess the
adiabatic condition for two-qubit operations in a double dot of
realistic dimensions.  In particular, we would like to determine the
minimum duration of two-qubit operations constrained by adiabaticity
and the quantum error correction threshold.  
%

To deal with a system described by a slowly-time-dependent
Hamiltonian, the adiabatic approximation can be employed, in which the
state of a quantum system at time $t$ can be expanded on the
instantaneous eigenstates at that time \cite{Schiff}: $\psi(t) =
\sum_i c_i(t) u_i(t)$ and $H(t) u_i(t) = E_i(t) u_i(t)$.  Substituting
these two equations into the time-dependent Schr\"{o}dinger equation,
the time evolution of the coefficients $c_i(t)$ can be obtained:
\begin{equation}
\frac{\partial c_k(t)}{\partial t} = \sum_{i \neq k}^{N}
\frac{c_i(t)}{E_k(t) - E_i(t)} \langle k |\frac{\partial
H(t)}{\partial t}| i\rangle \,  \exp\left\{\frac{1}{i\hbar}
\int_{-\infty}^t (E_i(\tau) - E_k(\tau)) d\tau  \right\} \,.
\label{eq:coef}
\end{equation}
In our calculation for the exchange gate of a QDQC, we consider a
two-electron double quantum dot \cite{HD}.  The basis states
$u_i$ are the two-electron eigenstates of the double dot.  The
explicit time-dependence is in the inter-dot barrier height.  In
other words, our complete confinement potential is given by  
$V({\bf r},t) = V_L({\bf r}) + V_R({\bf r}) + V_b ({\bf r},t)$,
where $V_L({\bf r})$ and $V_R({\bf r})$ are the fixed single quantum
dot confinement potentials, while $V_b ({\bf r},t)$ is a
Gaussian barrier potential that takes the form
$V_b({\bf r},t) = \left[ V_{max} - (V_{max} -
V_{min}) e^{-\left( \frac{t}{\tau} \right)^2} \right] e^{-\left(
x^2/l_x^2 + y^2/l_y^2 \right)}$.
Here $V_{max}$ and $V_{min}$ are the highest and lowest barriers
respectively, $l_x$ and $l_y$ defines the width and thickness
of the potential barrier between the two QDs located at
$x=\pm a$ on the $y$ axis, and $\tau$ is the half width of the 
Gaussian temporal variation of the barrier height.  We have chosen
to use Gaussian lineshapes for the confinements, barrier, and time
dependence without loss of generality.  Indeed, as we point out below,
non-Gaussian-shaped pulses might produce more favorable conditions
for the operation of two-qubit gates.  Now the matrix
elements of the time derivative of the Hamiltonian can be expressed
completely in terms of the potential barrier matrix elements:
$
\langle i| \frac{\partial H(t)}{\partial t} |j \rangle 
= \langle i| \frac{\partial V_b(t)}{\partial t} |j \rangle
= \left( \langle i| V_{max} |j \rangle 
- \langle i| V_b(t) |j \rangle \right) \frac{2t}{\tau^2} \,.
$
The wavefunction coefficients can then be calculated by plugging
these matrix elements into Eq.~(\ref{eq:coef}).

We perform a numerical calculation for the evolution of a two-electron
double dot using Eq.~(\ref{eq:coef}).  Initially the system is
entirely in the ground singlet state.  In other words, $c_i =
\delta_{i0}$ for all $i$.  The variation of the potential barrier will
not lead to singlet-triplet coupling.  Both types of states have about
the same excitation gap (smaller of the single particle excitation
energy  and the on-site Coulomb charging energy) and similar energy
spectra, and the exchange coupling does not lead to transitions
between singlet and triplet states, so that a pure singlet initial
state should give a reliable description of adiabaticity for a general
case.  By varying the half width $\tau$ of the barrier variation
time, we quantitatively evaluate the change in the ground singlet
state population, thus obtaining a lower limit to the gate operating
time using the criterion of quantum error correction code threshold. 
In the direct integration of Eq.~(\ref{eq:coef}), spectrum information
such as the barrier matrix elements and eigen-energies are needed.  We
do not attempt to calculate the state spectrum at all barrier
heights, as it would have made the whole computation intractable. 
Instead, we perform the spectrum calculation at a series of potential
barrier heights (at 10 points which correspond to 19 points on the
Gaussian temporal profile of the barrier variation), then obtain the
rest of the information through interpolation.  The energy spectra we
use are for a double dot with Gaussian confinements of 30 nm
radii and 40 nm inter-dot distance \cite{HD}.  The energy barrier
height $V_b$ ranges between 14 meV and 35 meV, corresponding to
exchange splitting of 280 $\mu$eV to 3.3 $\mu$eV.  

The result of our calculation is plotted in Fig.~1(a).  The leakage
(y-axis) is defined as $1-|c_0|^2$ which is zero before the gate is
applied, and should be zero if the gate is perfectly adiabatic. The
curve has several interesting features.  At short gating times the
leakage decreases quite fast, with the straight line in Fig.~1 giving
an indication of the speed of this decrease.   This decrease rate
slows down as the gating time increases.  At very long time, adiabatic
approximation should be asymptotically exact, so that the $k$th
excited state occupation takes the form
\begin{equation}
|a_k|^2 = \frac{4 \sin^2 \omega_{k0}t/2}{\hbar^2 \omega_{k0}^4}
\left|\langle k|\frac{\partial V_b}{\partial t}|0\rangle\right|^2 \,.
\end{equation}
The magnitude of the probability amplitude $a_k$ is in the order of
the ratio between the change in the Hamiltonian in one Bohr period 
$2\pi/\omega_{k0}$ and the energy difference $\hbar \omega_{k0}$.
Thus the leakage should be proportional to $1/T^2$ at large $T$
if we qualitatively represent $\partial H/\partial t$ by $\Delta
H/T$ where $T$ characterizes the total gating time.

The leakage we obtained
oscillates quite rapidly with the gating time.  This oscillation is
due to the phase factors $\exp\left\{\frac{i}{\hbar} \int_{-\infty}^t
\Delta E_{ki}(\tau) d\tau \right\}$ for all the levels involved.  As
the gating time is varied, the contribution to the leakage from
different higher levels changes rapidly, leading to the fast
oscillations.

Figure~1(a) demonstrates that for gating time ($\sim 2\tau$) longer than
30 ps, leakage in our double dot system should be
sufficiently small ($\lesssim 10^{-6}$) so that the currently
available quantum error correction schemes would be effective.  On the
other hand, an exchange splitting of 0.1 meV corresponds to about 20
ps gating time for a swap gate \cite{LD} (with rectangular pulse) at
the shortest.  Therefore, adiabatic condition does not place an extra
burden on the operation of the two-qubit gates such as a swap.  There
is in general no need to significantly increase the gating time in
order to accommodate the adiabatic requirement.

Figure~1(b) shows the energy splittings between the ground singlet
and the 1st and 2nd excited singlet states.  
A finite gap between the ground and excited states exists for all
barrier heights, thus guaranteeing the relatively easy satisfaction of
adiabatic condition in this configuration.  The anti-crossing in this
figure shows a competition between S-P hybrid states (one electron in
the single dot S states while the other in the single dot P states
\cite{HD}) and double occupied states (both electrons are in the S
state of the same dot) in the double dot.  Essentially, this
competition indicates the relative importance of single particle
excitation (S-P transition) and on-site Coulomb interaction.  At high
barrier height, the S-P hybrid state has lower energy than the double
occupied state, while at lower barrier the 1st excited singlet state
is a mixture of the double occupied state and the S-P hybrid state.
As the inter-dot barrier is lowered, the two electrons
can tunnel between the two dots more easily, thus reducing the effect
of on-site Coulomb repulsion.

The duration of the gate voltage pulse that changes the barrier height
between two neighboring QDs is not the only parameter that
determines whether adiabaticity is satisfied.  Pulse shape
tailoring is extensively used in optical control experiments
\cite{pulseshape}, and is presumably applicable in the
control of interdot barrier.  Our choice of a Gaussian pulse shape is
by no means optimal.  Therefore, the result shown here provides an
upper limit to the adiabatic condition for the adopted parameters in
the spin-based QDQC architecture. 

One condition for the adiabatic approximation to be valid is that the
system has no level crossing during the evolution.  In the double QD
system, there should be no orbital degeneracy as we
lower the inter-dot barrier.  The situation bears some similarity to
the transition from a hydrogen molecule to a helium atom,
although in the operation of a QDQC we would never push the two
QDs so close and the barrier so low that the two-electron double dot
becomes a two-electron single dot.  Here the state spectrum
essentially maintains the same structure throughout the gate
operation. It would, however, be interesting to explore the state
spectrum as the double dot is squeezed towards a single QD, and study
whether degeneracy might play a role during such evolution.  After
all, in a QDQC, QDs are used to label the qubits; while in a
two-electron QD, labeling is impossible as the two electrons are
indistinguishable.

Let us now turn our attention to the Cooper pair box quantum computer
(CPBQC) model.  The Hamiltonian of a CPB can be written on the basis
of charge number states of the box:
\begin{eqnarray}
H & = & 4E_C (\hat{n}-n_g)^2 -E_J \cos\hat{\phi} \nonumber \\
& = & \sum_n \left[ 4E_C (n-n_g)^2 |n\rangle \langle n| -
\frac{E_J}{2}(|n\rangle\langle n+1| + |n+1\rangle\langle n|) \right]
\,, 
\label{eq:CPB}
\end{eqnarray}
where $E_C$ is the charging energy of a Cooper pair box (CPB), $E_J$
is the Josephson coupling between the CPB and an external
superconducting lead, $n_g$ represents the applied voltage on the
CPB in terms of an effective charge number, and $n$ refers to the
number of Cooper pairs in the box.  The stationary solutions of a CPB
form energy bands.  The two levels for a CPB qubit are the lowest two
energy levels at $n_g=1/2$, where the eigenstates are approximately
$|\!\downarrow\rangle=(|0\rangle + |1\rangle)/\sqrt{2}$ and
$|\!\uparrow\rangle=(|0\rangle - |1\rangle)/\sqrt{2}$ with a splitting
of about $E_J$.

Most theoretical studies of CPBQC employ a two-level description of
the system \cite{Schon}.  However, as we have learned \cite{HD,HD2}
in the case of QDQC above, higher excited states can play an
important role in the dynamics of a qubit, especially when it is
subjected to non-resonant operations.  One particular operation for
CPBQC is the sudden pulse gate to shift $n_g$, thus bringing a system
from a pure ground state ($|0\rangle$ at, e.g. $n_g=1/4$) to a
coherent superpositioned state ($(|\!\uparrow\rangle +
|\!\downarrow\rangle)/\sqrt{2}$ at $n_g=1/2$).  However, such a simple
description of the pulse gate is only valid when $E_J/E_C \rightarrow
0$.  Since $E_J$ determines the gate speed of a CPBQC, such a
condition is not practical for a realistic QC.  Below we explore the
situation when $E_J$ is not vanishingly small, and study how the pulse
gate operation will be affected.

Before considering the pulse gate, we first study the 
state compositions of a CPB in order to find the limit of a two-level
model and show some analytical results in Table~\ref{table1} (For
these expressions, we include only the lowest excited states). 
As is shown in Table~\ref{table1}, no matter which $n_g$ value we
choose, there are always unwanted states that mix into the composition
of the eigenstates with weights proportional to $r=E_J/E_C$.  Such
mixtures will affect the dynamics of the system.  For example, if we
prepare a CPB in the ground state $\psi_0$ at $n_g=1/4$ and apply a
sudden pulse gate to shift $n_g$ up to $1/2$, the state right after
the shift will not be $(|\!\uparrow\rangle +
|\!\downarrow\rangle)/\sqrt{2}$, but $\approx (|\!\uparrow\rangle +
|\!\downarrow\rangle)/\sqrt{2} + \frac{r}{4}(|\!\uparrow\rangle -
|\!\downarrow\rangle)/\sqrt{2}$.  Thus, finite $r$ will lead to
errors in the pulse gate to the first order of $r$. 

In real experiments, the pulse gate always has finite rise/fall times
(non-sudden).  In Ref.~\onlinecite{Nakamura}, the pulse rise time is
in the range of 30 to 40 ps.  Such gradual rise and fall of the
pulse gate inevitably lead to more errors, which have been considered
in the context of two-level systems \cite{Choi,Oh}.  Here we
calculate the fidelity of the pulse gate taking into account the
finite rise/fall time, the higher excited states, and the complete
composition of all the eigenstates.  The eigenstates are
numerically obtained (instead of perturbatively \cite{Wang}) on the
basis of 21 charge number states ($|-10\rangle$, $\cdots |0\rangle$,
$|1\rangle$, $\cdots |10\rangle$). We prepare the initial state as
the ground state $\psi_0$ near $n_g=1/4$, then let the CPB evolve
under Hamiltonian (\ref{eq:CPB}).  The pulse gate represented by
$n_g$ is turned on with a lineshape ranging from linear to quintic
power of time during the rise and fall: $n_g(t)=n_g(0) +
\left(\frac{2t}{t_r} \right)^m (n_g(t_r)-n_g(0))/2$ for $t \in
[0,t_r/2]$ and $n_g(t)=n_g(t_r) - \left[\frac{2(t_r-t)}{t_r}
\right]^m (n_g(t_r)-n_g(0))/2$ for $t \in [t_r/2,t_r]$, where $t_r$
is the pulse rise/fall time and the pulse is switched on from $0$ to
$t_r$.  For system parameters, we choose the values used in
Ref.~\onlinecite{Nakamura}, with $E_C = 117$ $\mu$eV, $E_J=51.8$
$\mu$eV, $n_g(0)=0.255$, and $n_g(t_r)=0.5$.

In Fig.~\ref{fig2} we plot the state fidelity as a function of rise
time.  Here the state fidelity is defined as the maximum probability
for the CPB to be in the first excited state after the pulse gate
when $n_g$ returns to $0.255$.  Figure~\ref{fig2} illustrates two
important points.  First of all, the fidelity of the pulse gate
decreases oscillatorily instead of monotonously as the pulse rise time
increases.  The oscillations (with periods of 30 to 40 ps) in the
curves represent the coherent evolution of the CPB during the
rise/fall of the pulse voltage: at $n_g=1/4$, the dominant energy
scale is $2E_C \approx 234 \ \mu$eV with a corresponding period of 17
ps, while at $n_g = 1/2$ the dominant energy scale is $E_J = 51.8 \
\mu$eV with a corresponding oscillation period of 76 ps.  The
observed oscillation period is essentially an average of these
two limits.  Secondly, including higher
excited states is important in correctly evaluating the fidelity
dependence on pulse rise time, as a comparison of the two panels of
Fig.~\ref{fig2} clearly demonstrates, especially for higher power of
time-dependence.  The real pulse shape may not be the same as any of
the power law time-dependence we consider here, but the disagreement
between the two panels in Fig.~\ref{fig2} clearly shows again
the necessity of taking excited states into consideration when
evaluating non-resonant gates such the pulse gate here.

Our current study is in essence a quantitative delineation of the
Hilbert space for a QDQC (CPBQC) during a two-qubit (single-qubit)
operation.  Such a study is necessary for other proposed QC schemes as
well in order to select the optimal operating regime, eliminate
possible errors, and enhance signal to noise ratios in experiments. 
We have restricted ourselves to $T=0$ in our calculations, but our
results remain valid at finite (low) temperatures as long as
substantial thermal occupation of higher excited levels does not
occur in the system. 

We thank David DiVincenzo and Belita Koiller for useful discussions,
and ARDA and LPS for financial support.

\begin{figure}
\centerline{
\epsfxsize=5.0in
\epsfbox{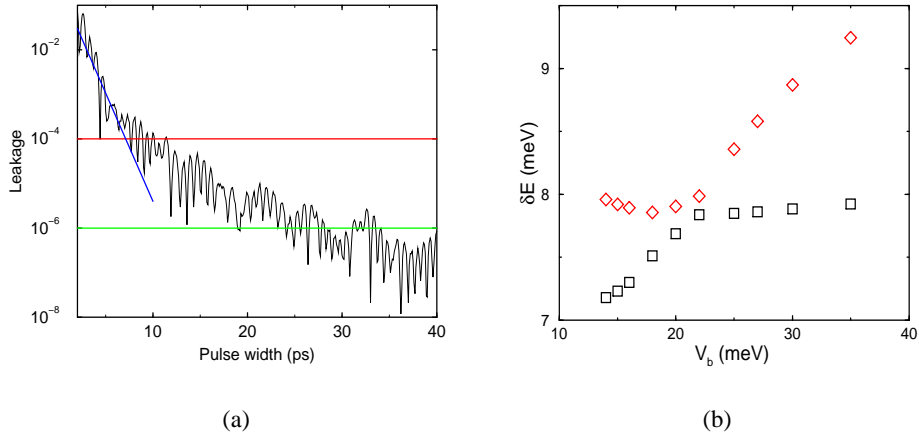}}
\vspace*{0.1in}
\protect\caption[]
{\sloppy{
In (a) we plot the state leakage as a function of the pulse width
$2\tau$ of the exchange gate.  The two-electron state is initially
prepared in the ground singlet state.  The two horizontal lines
represents the commonly used threshold for quantum error correction. 
The fitted line at the small pulse-width indicates the initial
rapid decrease in the error rate (leakage) as the pulse becomes
wider (or the gate operation becomes slower).  In (b) we plot the 
energy splitting between the ground singlet state and the first and
second excited singlet states as functions of inter-dot barrier
height.  The presence of this gap for all barrier heights is
important in making the adiabatic condition easier to satisfy.
}}
\label{fig1}
\end{figure}
\begin{figure}
\centerline{
\epsfxsize=5.0in
\epsfbox{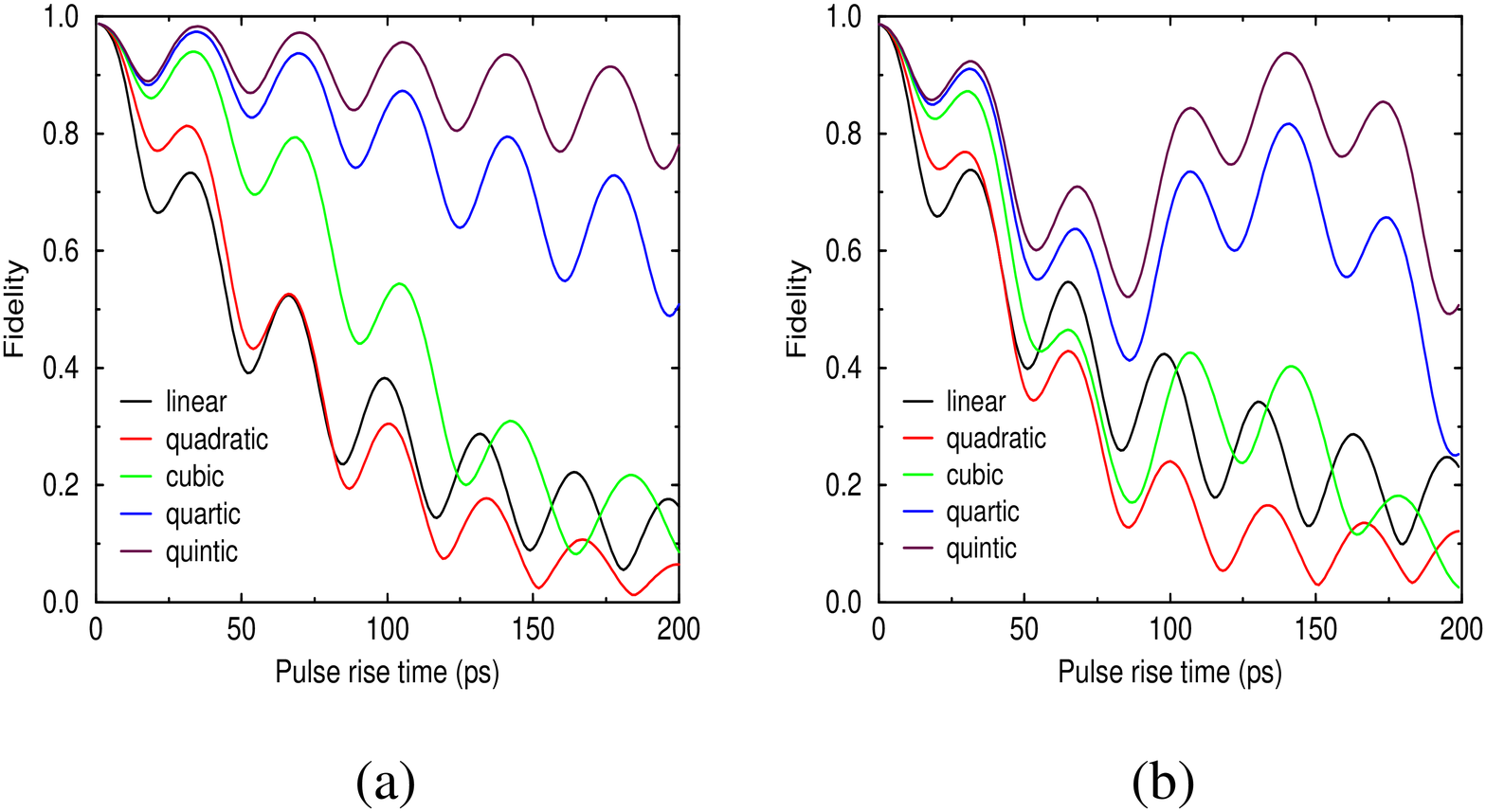}}
\vspace*{0.1in}
\protect\caption[]
{\sloppy{
State fidelity as a function of the finite rise/fall time of a pulse
gate in a single Cooper pair box.  The fidelity here is defined as
the maximum probability of the CPB in the first excited state after
the application of a pulse gate with the state starting from the
ground state (represented by $n_g$ goes from 0.255
to 0.5, then back to 0.255 after a period of time $\tau_p$).  In panel
(a), the CPB is treated as a two-level system (on the basis of
$|0\rangle$ and $|1\rangle$), while in panel (b) the CPB is treated
as a multi-level system (on the basis of $|-10\rangle, \cdots, 
|0\rangle, |1\rangle, \cdots, |10\rangle$).The five curves in each
panel represent the different lineshapes of the rise/fall of the
pulse ranging from linear to quintic, as described in the text. 
}} 
\label{fig2}
\end{figure}
%
%
\begin{table}
\caption[]{Eigenstates and energies of a Cooper pair box.}  
\begin{center}
\begin{tabular}{cccc}  
$n_g$ & low energy eigenstates & $b/a$ ($r=E_J/E_C$) & 
eigenenergies \\  \hline
0 & $\psi_0 = a|0\rangle + b(|1\rangle+|-1\rangle) $ & $r/8$ & 
-$\frac{E_J^2}{8E_C}$  \\ 
& $\psi_1 = (|1\rangle-|-1\rangle)/\sqrt{2}$ & --- & $4E_C$ \\
& $\psi_1 = a(|1\rangle-|-1\rangle) - b|0\rangle$ & r/4 & 
$4E_C + \frac{E_J^2}{8E_C}$ \\
$\approx 1/4$ & $\psi_0 = a|0\rangle + b[\frac{|-1\rangle}{1+2n_g}
+ \frac{|1\rangle}{1-2n_g}]$ & $r/8$ & $4E_C n_g^2 -
\frac{E_J^2}{8E_C (1-4n_g^2)}$  \\   
& $\psi_1 = a|1\rangle + b[\frac{|2\rangle}{3-2n_g}
+ \frac{|0\rangle}{1-2n_g}]$ & $r/8$ & $4E_C (1-n_g^2) -
\frac{E_J^2}{8E_C (3-2n_g^2)(1-2n_g^2)}$  \\  
$1/2$  & $\psi_0 = a(|0\rangle + |1\rangle)+b(|-1\rangle +
|2\rangle)$ & $r/18$ & $E_C - E_J/2 -\frac{E_J^2}{32E_C}$   \\
& $\psi_1 = a(|0\rangle - |1\rangle)+b(|-1\rangle -
|2\rangle)$ & $r/18$ & $E_C + E_J/2 -\frac{E_J^2}{32E_C}$   \\
\end{tabular}
\end{center}
\label{table1}
\end{table}


\begin{thebibliography}{99}
\bibitem{Multi} We have shown that under certain conditions
multielectron QD can also be used as qubit for a QDQC \cite{HD2}. 
The considerations regarding adiabatic condition would be similar in
that architecture, while here we will focus on single-electron QDs.
\end{thebibliography}
\end{document}